\documentclass{article}
\usepackage{spconf,amsmath,graphicx,url,times}
\usepackage{hyperref}
\newcommand*{\email}[1]{\href{mailto:#1}{\nolinkurl{#1}} }
\usepackage{microtype}
\usepackage{booktabs}
\usepackage{color}
\newcommand{\citep}[1]{\cite{#1}}
\newcommand{\citet}[1]{\cite{#1}}
\usepackage{amsmath}
\DeclareMathOperator*{\argmin}{arg\,min}
\usepackage{amssymb}
\newcommand\norm[1]{\left\lVert#1\right\rVert}

\title{Mask-dependent Phase Estimation for Monaural Speaker Separation}

\twoauthors
 {Zhaoheng Ni}
	{The Graduate Center, CUNY\\
	New York, NY, USA\\
	\email{zni@gradcenter.cuny.edu}
	}
 {Michael I Mandel}
	{Brooklyn College, CUNY\\
	Brooklyn, NY, USA\\
	\email{mim@sci.brooklyn.cuny.edu}
	}
\begin{document}

\ninept
\maketitle

\begin{sloppy}

\begin{abstract}
    Speaker separation refers to isolating speech of interest in a multi-talker environment. Most methods apply real-valued Time-Frequency (T-F) masks to the mixture Short-Time Fourier Transform (STFT) to reconstruct the clean speech. Hence there is an unavoidable mismatch between the phase of the reconstruction and the original phase of the clean speech. In this paper, we propose a simple yet effective phase estimation network that predicts the phase of the clean speech based on a T-F mask predicted by a chimera++ network. To overcome the label-permutation problem for both the T-F mask and the phase, we propose a mask-dependent permutation invariant training (PIT) criterion to select the phase signal based on the loss from the T-F mask prediction. We also propose an Inverse Mask Weighted Loss Function for phase prediction to focus the model on the T-F regions in which the phase is more difficult to predict. Results on the WSJ0-2mix dataset show that the phase estimation network achieves comparable performance to models that use iterative phase reconstruction or end-to-end time-domain loss functions, but in a more straightforward manner.
\end{abstract}

\begin{keywords}
Speech separation, phase estimation, permutation invariant training, chimera++, deep learning 
\end{keywords}

\section{Introduction}
\label{sec:intro}

Recently, many deep-learning approaches have been proposed for the monaural multi-talker speech separation problem. \cite{hershey2016deep} proposed a deep clustering method that projects each time-frequency bin to a high-dimensional vector space using a bi-direction long short-term memory (BLSTM) network. T-F masks can be generated by clustering these vectors using the k-means algorithm. While deep clustering can separate mixtures when the number of speakers is unknown without changing the model architecture, it produces binary time-frequency masks, which may hurt the intelligibility of the speech. \cite{erdogan2015phase} showed that phase-sensitive masks perform better than ideal binary masks on the speech enhancement task. To overcome the label-permutation problem in multi-talker speech separation, \cite{yu2017permutation} proposed the permutation invariant training (PIT) criterion which lets the deep neural network select the assignment of output sources to ground truth sources using the lowest loss value among all permutations. \cite{luo2017deep} combined deep clustering with a mask-estimation model in a single hybrid network, called the chimera network. Their results show that adding deep clustering leads to better mask predictions, thus achieving better separation performance. \cite{wang2018alternative} made slight modifications to the original chimera network and developed several alternative loss functions for both deep clustering and mask-inference outputs. The new network (dubbed ``chimera++'') boosted separation performance using a much simpler architecture. 

While there have also been many recent developments in mask-based separation approaches, there are two main issues remaining to be solved:
\begin{itemize}
    \item Since most masks are restricted to be between 0 and 1, there will be an unavoidable error if the magnitude of the clean speech is greater than that of the mixture due to phase cancellation.
    \item Most of these methods utilize the phase of the mixture as that of the separated speech, which causes an unavoidable phase difference error.
\end{itemize}
The phase-sensitive mask (PSM) was introduced~\cite{erdogan2015phase} to decrease the impact of such phase differences. The PSM as a mask prediction target is defined as:
\begin{equation}
    M_{\text{PSM}} = \frac{|S|}{|X|} \odot \cos{(\angle{S}-\angle{X})},
\end{equation}
where $S$ and $X$ are the STFT of the clean speech and the speech mixture respectively at any particular time-frequency point, $\odot$ is the element-wise multiplication. By applying the PSM to the mixture, the real component of the estimated STFT is close to that of the ground truth. 

Besides the PSM, \citep{williamson2016complex} introduced the complex ratio mask (cRM) to enhance both the magnitude and the phase spectra by operating in the complex domain. The real component of the cRM is equivalent to the PSM. However, \cite{williamson2017speech} showed that the imaginary component is difficult to predict directly from the noisy magnitude due to the randomness of the phase pattern.

To estimate the phase of each separated source, \cite{wang2018alternative} started with the estimated magnitude and the noisy phase and jointly reconstructed the phase of all sources using the multiple-input spectrogram inversion (MISI) algorithm \citep{gunawan2010iterative}. Later, \cite{wang2018end} unfolded the iterations in the MISI algorithm into separate STFT and iSTFT layers on top of a chimera++ network and trained the network to minimize the error in the time domain. Their results showed that it is possible to predict the phase of separated sources from a magnitude estimate and the noisy phase. 

Instead of using an iterative STFT and iSTFT to reconstruct the phase, the current paper proposes an approach to reconstruct the phase directly from the magnitude estimate and the noisy phase. The problem can thus be formulated as estimating the magnitude and phase of each source from the noisy STFT, $X$
\begin{equation}
\label{eq:formulation}
    S_c = |X| \odot M_c \odot \cos{\theta_c} + j |X| \odot M_c \odot \sin{\theta_c},
\end{equation}
where $|X|$ is the noisy magnitude, $M_c$ and $\theta_c$ are the mask and phase estimates for source $c$, respectively. 

Some studies have tried to predict the mask and phase directly without iterative algorithms. In \citep{ephrat2018looking} and \citep{afouras2018conversation}, the authors utilized visual feature to drive the model to predict the corresponding mask and phase. However, some datasets may not contain visual data, or the video may not capture the speaker of interest. Though the PIT criterion solves the label-permutation problem for mask estimates, integrating PIT with both mask and phase estimates is still an open problem. Thus, it is still essential to figure out how to apply PIT criterion in phase estimation with only the audio data.

Another problem in phase estimation is that its difficulty varies across regions. Since the noisy STFT $X = \sum_c{S_c}$, where $S_c$ represents the clean STFT for source $c$, if a T-F bin is only dominated by one speaker (i.e., the mask value is close to one), the phase difference between noisy and clean is close to zero.  In regions where the mask is not close to one, the phase difference is influenced by the magnitude and phase of all significant sources, and is thus more difficult to estimate.

Hence this study proposes a joint-training algorithm that estimates the ideal ratio mask (IRM) by using a chimera++ network and then estimates the clean phase based on the mask estimate and the noisy phase for each source. The label permutation problem is solved by using a PIT criterion based only on the estimated masks, with the phase estimated after this matching. Three different weighed loss functions are proposed to compare the influence of different T-F regions over the training of the phase estimation model.

\section{Chimera++ Network}
\label{sec:chimera++}


This section describes the architecture of the chimera++ network and its loss function, which is also incorporated into our algorithm.  The left half of Figure~\ref{fig:phase-net} shows the architecture of the chimera++ network. Each time-frequency bin is projected to a D-dimensional vector $v_i \in \mathbb{R}^{1\times D}$ via a deep clustering layer\citep{hershey2016deep}, where $i$ corresponds to a particular pair of time and frequency indices. Each T-F bin has a one-hot label vector $y_i \in \mathbb{R}^{1 \times C}$ indicating the speaker among $C$ speakers who dominates this bin. Stacking all embedding vectors and labels produces the embedding matrix $V \in R^{TF\times D}$ and the label matrix $Y \in R^{TF\times C}$. The objective of deep clustering is to group together the embedding vectors of T-F bins from the same speaker and make those from different speakers orthogonal. The loss function is defined as:
\begin{eqnarray}
    L_{\text{DC, classic}} &= &\norm{VV^T - YY^T }_F^2  
\end{eqnarray}
\cite{wang2018alternative} showed reducing the influence of silence T-F bins benefits training the deep clustering network. They introduced binary voice activity weights $W_{\text{VA}}$ to the loss function:
 \begin{eqnarray}
     L_{\text{DC, classic, W}} &= & \norm{W_{\text{VA}}^{\frac{1}{2}}(VV^T - YY^T)W_{\text{VA}}^{\frac{1}{2}}}_F^2  
 \end{eqnarray}

In terms of the loss function for the mask inference layer, \cite{wang2018alternative} recommends using a loss function based on the truncated phase-sensitive spectrum approximation (tPSA), defined as:
\begin{multline}
     L_{\text{MI}, \text{tPSA}} = \min_{\pi \in P}\sum_{c} \left\| \hat{M_{c}}\odot \left| X\right| \right. \\ \left. -
     T_0^{\left| X\right|}(\left| S_{\pi(c)}\right| \odot \cos (\theta_{X} - \theta_{\pi(c)})) \right\|_F^{1},
\end{multline}
where $P$ is the set of permutations on $\{1,\dots,C\}$, $X$ and $S$ are the STFT of the noisy and clean speech respectively, $T$ is the truncation function defined as $T_a^b(x) = \min(\max(x,a),b)$, $\theta_X$ and $\theta_c$ are the noisy phase and the true phase of source $c$, respectively. 

However, estimating tPSA contradicts with estimating the phase vectors in our problem formulation since both consider the cosine of the clean phase. To adapt the estimate into our problem formulation, we change the tPSA loss function to the magnitude spectrum approximation (MSA) for magnitude reconstruction.
\begin{equation}
     L_{\text{MI}, \text{MSA}} = \min_{\pi \in P}\sum_{c}{\norm{\hat{M_{c}}\odot \left| X\right| -
     \left| S_{\pi(c)}\right|}_F^{1}}.
\end{equation}

We compare our chimera++ implementation with that in \citep{wang2018alternative} in Table~\ref{tab:chimeraComparison}, which shows that we have successfully reproduced the reported result. We find that our implementation achieves slightly better performance by using the classic loss function based on tPSA, compared that using a whitened k-means loss function (W) \citep{wang2018alternative} based on tPSA. This indicates we can fairly compare our phase estimation network with the other methods which are also based on the chimera++ network.

\begin{table}
    \centering
    \begin{tabular}{llcc}
    \toprule
    Model & Loss Function & Mask Type & SDR\\
    \midrule
    Chimera++ \citep{wang2018alternative}& $\textrm{DC}(W, W_{\textrm{VA}})$ & tPSA & 10.9\\
    Our implementation & $\textrm{DC}(\textrm{classic}, W_{\textrm{VA}})$ & tPSA & 11.0\\
    \bottomrule
    \end{tabular}
    \caption{Comparison between the chimera++ in \cite{wang2018alternative} and our implementation}
    \label{tab:chimeraComparison}
\end{table}

\section{Phase Estimation Network}
\label{sec:phase_network}

\begin{figure}
  \centering
  \centerline{\includegraphics[width=\columnwidth]{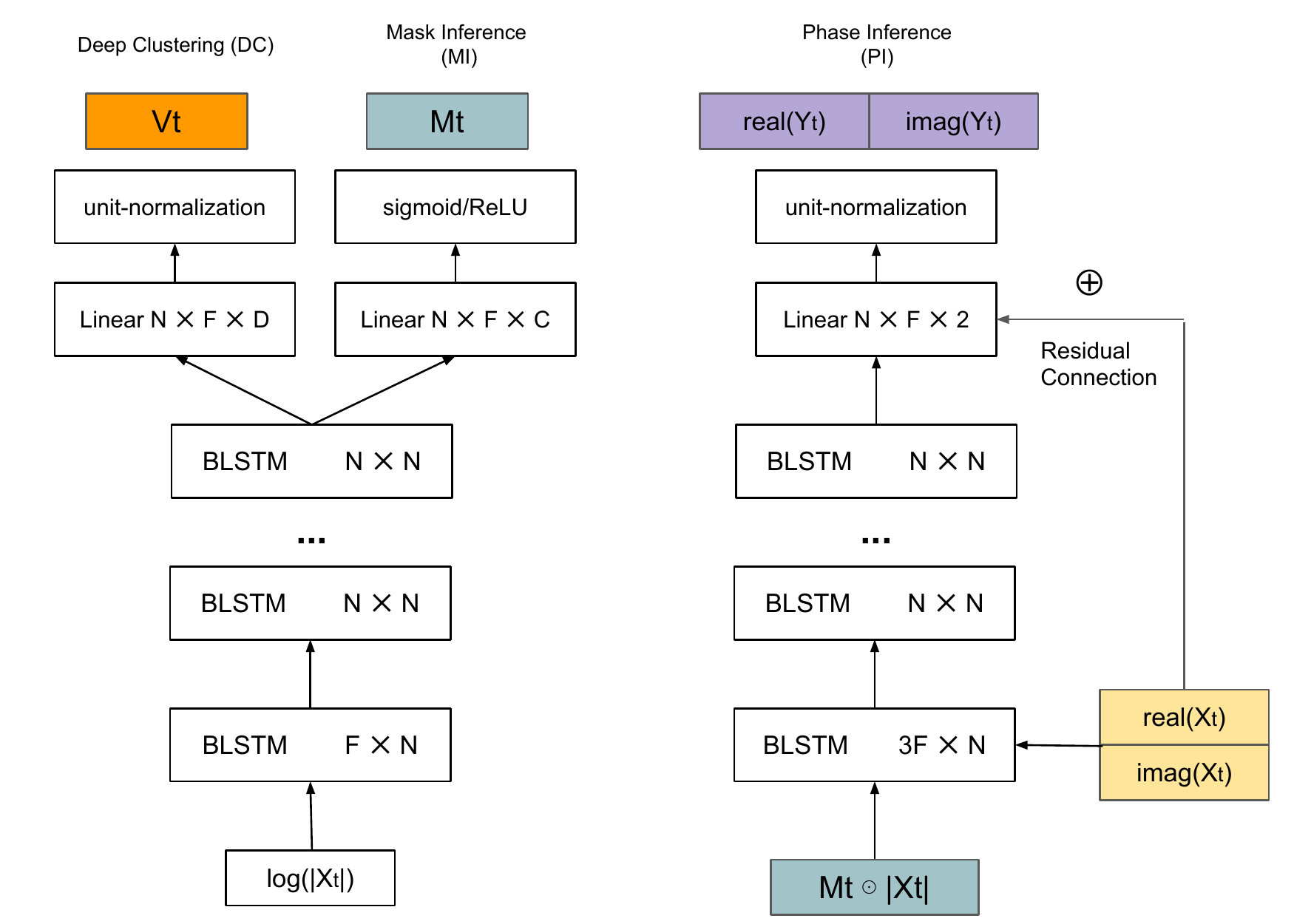}}
  \caption{Model architecture of the proposed phase estimation network. Note that the left tower is the original chimera++ network.}
  \label{fig:phase-net}
\end{figure}

Figure~\ref{fig:phase-net} shows the architecture of the proposed phase network. First the magnitudes of each of the sources are estimated by applying the masks estimated by the chimera++ network to the noisy magnitude. Each magnitude estimate is concatenated with the real and imaginary components of the noisy STFT as the input feature for the phase inference (PI) sub-network. The output is a $T\times F\times 2$ matrix representing the cosine and sine of the phase estimate for each T-F bin. Each phase estimate is generated based on one of the masks. The output of the final linear layer is added to the original real and imaginary components of the noisy phase as a residual connection~\cite{he2016deep} before unit-normalization. The loss function for this prediction is 
\begin{equation}
    L_{\text{PI}} = -\sum_{c,t,f}{ \langle \hat{p}_{t,f}^c, p_{t,f}^c \rangle}
\end{equation}
where $p_{t,f}^c$ is the vector $\langle \cos{\theta_c}, \sin{\theta_c} \rangle$ representing the true phase of the $t,f$ bin in source $c$ and $\hat{p}_{t,f}^c$ is the corresponding phase estimate.

\subsection{Mask-dependent PIT Criterion}
\label{subsec:mask-dependent-loss}

The PIT criterion~\citep{yu2017permutation} has been successful in solving the label-permutation problem.  To overcome the permutation problem for both the mask and the phase, several combinations of mask and phases losses could be used for selecting the best permutation. Since the phase estimate is dependent on the mask, the simplest combined PIT is based on a combination of both the losses of the MI and PI layers together.
We use $L_{\text{PIT}}$ to represent this combined loss
\begin{multline}
    L_{\text{PIT, MP}} = \min_{\pi \in P}\sum_{c} \Big( \norm{\hat{M_{c}}\odot \left| X\right| - \left| S_{\pi(c)}\right|}_F^{1} \\
       -  \sum_{t,f} \langle \hat{p}_{t,f}^c, p_{t,f}^{\pi(c)} \rangle \Big)
\end{multline}
But since the phase is more difficult to learn to predict than the mask, perhaps because of a larger number of local minima, the phase tends to be misleading to this permutation matching in early epochs.  Instead we only utilize the mask for matching, and continue this throughout training. Hence we propose a Mask-dependent PIT criterion defined as
\begin{eqnarray}
   L_{\text{PIT, MD}} &= &\sum_{c}{\norm{\hat{M_{c}}\odot \left| X\right| -
     \left| S_{\pi(c)}\right|}_F^{1}} - \sum_{c,t,f}{ \langle \hat{p}_{t,f}^c, p_{t,f}^{\pi(c)} \rangle}, \nonumber \\
   &&\pi =\argmin_{\pi}{\sum_{c}{\norm{\hat{M_{c}}\odot \left| X\right| -
     \left| S_{\pi(c)}\right|}_F^{1}}}.
\end{eqnarray}

\subsection{Weighted Loss Functions for Phase Estimation}

Phase is harder to predict in certain T-F regions. Figure~\ref{fig:phase-visual} shows for one source in a mixture the log magnitude of the clean speech, IRM, $\cos$, and $\sin$ of the phase difference $\angle\theta_{\text{S}} - \angle\theta_{\text{X}}$. If the mask value is close to one, meaning the phase of the mixture mostly comes from this source, the $\cos$ of the phase difference is close to 1 and the $\sin$ of the phase difference is close to 0. Hence the phase difference between the clean and the noisy STFT is very small. When the mask value is small, the phase differences are larger, making it more challenging to predict the phase.

\begin{figure}
  \centering
  \centerline{\includegraphics[width=\columnwidth]{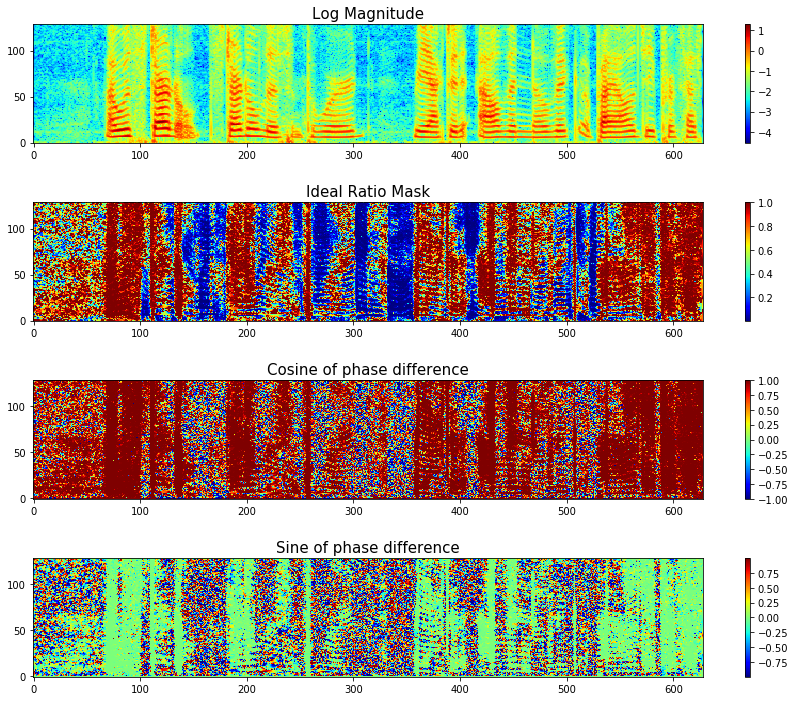}}
  \caption{The clean spectrogram, the IRM, cosine, and sine of phase difference between clean and noisy STFT (sample utterance: \url{cv/s2(mix)/011a010g_0.16366_40pc0204_-0.16366.wav})}
  \label{fig:phase-visual}
\end{figure}

In the MI loss of the chimera++ network, an MSA or tPSA loss function is introduced to increase the weight of the losses where the energy of the noisy magnitude is higher. However, the same weight may not be suitable for estimating the phase. We propose three different weighted loss function for the PI loss to differently weight the contributions to the phase estimates.

We define the Magnitude Weighted Loss Function (MWL) as
\begin{equation}
    L_{\text{PI, MWL}} = -\sum_{c,t,f}{(\gamma + M_{c, t,f}) \langle \hat{p}_{t,f}^c, p_{t,f}^{\pi(c)} \rangle},
\end{equation}
where $M_{c, t,f}$ represents the magnitude of source $c$ on the $t,f$ bin, $\pi$ is identified by the PIT loss, and $\gamma$ is a tunable parameter that avoids applying 0 loss to points with 0 magnitude.  We set it to 0.2 for our experiments. The motivation for this is that higher magnitude values should be emphasized because they are more reliable in training the phase network.

We define the Inverse Magnitude Weighted Loss Function (I-MWL) as
\begin{equation}
    L_{\text{PI, I-MWL}} = -\sum_{c,t,f}{(\gamma + M_{\neg c,t,f})\langle \hat{p}_{t,f}^c, p_{t,f}^{\pi(c)} \rangle},
\end{equation}
where $M_{\neg c,t,f} = \sum_{i \neq c}^C M_{i, t,f}$ and $\pi$ is identified by the PIT loss. The motivation for this is that lower mask values are more difficult to predict, so should be emphasized.

We define the Joint Weighted Loss Function (Joint) as
\begin{equation}
    L_{\text{PI, Joint}} = -\sum_{c,t,f} \langle \hat{p}_{t,f}^c, p_{t,f}^{\pi(c)} \rangle \sum_i M_{i,t,f}.
\end{equation}
The motivation for this weighting is to emphasize all active regions in the spectrogram. Only silent regions are ignored, similarly to the loss function of the deep clustering layer. Note that if the noisy magnitude is small while there are other sources whose magnitudes are large, the bin is emphasized more.

The loss function for the whole network thus becomes
\begin{equation}
    L_{\text{Phase Network}} = \alpha L_{\text{DC}} + (1-\alpha) L_{\text{PIT}},
\end{equation}
where $\alpha$ is a weight parameter that balances the loss of the DC layers and the MI and PI layers. We use 0.975 for $\alpha$, following \citep{wang2018alternative}.

\section{Experiments}

To evaluate our proposed method, the models are trained and validated on the WSJ0-2mix dataset~\citep{hershey2016deep}, which is a publicly available dataset for multi-talker speech separation. There are 20,000 mixtures ($\sim$30 hours) in the training data, 5,000 mixtures ($\sim$10 hours) in the validation data, and 3,000 mixtures ($\sim$5 hours) in the test data. The training data and the validation data are generated by randomly mixing two utterances of two random speakers in the WSJ0 training data (si\_tr\_s). Some speakers in the validation data are included in the training data with different utterances, thus the validation data is considered to be the closed speaker condition (CSC). The test data is generated by mixing two random  utterances of two random speakers from the WSJ0 test data (si\_et\_05). None of the speakers in the test data are included in either the training or the validation data, thus it is considered to be the open speaker condition (OSC). The sampling rate is 8~kHz and the Signal to Noise Ratio (SNR) of utterances is in a range between 0 and 10~dB.

The chimera++ network and the proposed phase network are implemented in PyTorch 1.0\footnote{https://github.com/speechLabBcCuny/onssen} \cite{paszke2017automatic}. Both the chimera++ network and the phase network have 4 BLSTM layers, each with 600 neural units in each direction. A 0.3 dropout rate is applied to the outputs of the BLSTM layers. Random chunks of 400 consecutive frames from each utterance are extracted to train the model with a batch size of 16 such chunks. The STFT is calculated by the librosa library~\cite{mcfee2015librosa} utilizing a Hann window. The FFT window size is 256 samples (32~ms) and the hop length is 64 samples (8~ms). The Adam optimizer with a $10^{-3}$ initial learning rate is used. Training is stopped when the validation loss does not decrease for 10 epochs. The model is trained for a total of 100 epochs if no early stopping mechanism is activated.

To evaluate the separation performance, the Signal to Distortion Ratio (SDR) computed with the mir\_eval library~\cite{raffel2014mir_eval} as the major evaluation metric. The baseline method is the noisy phase with the MSA magnitude estimated by the chimera++ network.

\begin{table}
    \centering
    \begin{tabular}{lcc}
    \toprule
        Method &  SDR & SI-SDR\\
        \midrule
        Chimera++, MSA & 10.5 & - \\
        ~~~~~~~~~+ tPSA \citep{wang2018alternative} & 11.5 & 11.2 \\
        ~~~~~~~~~+ MISI-5 \citep{wang2018alternative} & 11.8 & 11.5 \\
        ~~~~~~~~~+ WA-MISI-5 \citep{wang2018end} & 12.9 & 12.6 \\
        Phasebook, MISI-0 \citep{le2019phasebook} & - & 12.6\\
        ~~~~~~~~~+ MISI-5 \citep{le2019phasebook} & - & 12.8\\
        Chimera++(Encoder-BLSTM-Decoder) \citep{wang2019deep} &- & 11.9\\
        Sign prediction network \citep{wang2019deep}& 15.6 & 15.3\\
        \bottomrule
    \end{tabular}
    \caption{Published SDR/SI-SDR improvements of different phase estimation methods on the open speaker condition (OSC) of the WSJ0-2mix dataset.}
    \label{tab:sdr-baseline}
\end{table}

\begin{table}
    \centering
    \begin{tabular}{lllc}
    \toprule
    PIT Criterion & Mask Activation & Phase Loss & SDR\\
    \midrule
     MP & ReLU & MWL & 11.0\\
     MP & Sigmoid & MWL & 11.5\\
     MD & Sigmoid & I-MWL & 12.0\\
     MD & Sigmoid & MWL & 12.6\\
     MD & Sigmoid & Joint & 13.0\\
     MD & Sigmoid & Joint,$\alpha = 0.5$ & 13.6\\
     \bottomrule
    \end{tabular}
    \caption{SDR improvements of the proposed method with different settings on the OSC of the WSJ0-2mix dataset. PIT criteria Mask+Phase (MP) and Mask-dependent (MD). Phase losses magnitude-weighted loss (MWL), inverse magnitude-weighted loss (I-MWL), and joint weighted loss (Joint).}
    \label{tab:sdr-propose}
\end{table}

\section{Results}
\label{sec:results}

Tables~\ref{tab:sdr-baseline} and \ref{tab:sdr-propose} show the SDR or scale-invariant SDR (SI-SDR)~\citep{le2018sdr}  or both improvements of recently published methods and those of the proposed phase estimation networks, respectively. The cells are left blank if they are not reported in the publications. The chimera++ network with MSA loss achieves 10.5~dB SDR. Adding our phase estimation network with $\text{PIT}_{\text{MP}}$ criterion improves the performances by 1.0~dB, which is the same as that of the chimera++ network with the tPSA loss. By comparing the activation functions of the MI layer, ReLU does not improve the performance over Sigmoid. 

Comparing rows 1 and 3 of Table~\ref{tab:sdr-propose}, using the mask-dependent $\text{PIT}_{\text{MD}}$ criterion achieves 12.6~dB SDR, which is 1.1~dB higher than using the $\text{PIT}_{\text{MP}}$ criterion. This shows that the loss for the mask estimate is more reliable and stable than combining it with the phase loss for the PIT criterion. In terms of different weighted loss functions, the weight of all magnitudes (Joint) achieves the best result of 13.0~dB SDR, followed by T-F regions of the clean speech (MWL) and then noisy speech (I-MWL). Inspired by \cite{isik2016single}, we also apply a curriculum training strategy \citep{bengio2009curriculum} that trains the network using $\alpha=0.975$ first, then retrains the model using $\alpha=0.5$, increasing the performance to 13.6~dB.

Comparing our results with published phase-based methods in Table~\ref{tab:sdr-baseline}, our result is better than the chimera++ with 5 iterations of the MISI algorithm (MISI-5). We get competitive result to the chimera++ with 5 MISI iterations and an end-to-end waveform approximate loss function (WA-MISI-5) without applying the curriculum training strategy. This indicates the proposed network can predict phase directly from the mask generated by chimera++. We do not currently use the waveform approximate loss function in our model, but our mask-dependent PIT criterion is applicable to the such losses. Instead of choosing the minimum waveform difference, we choose the permutation that gives the minimum mask loss. The model still backpropagates to minimize the waveform difference, while the mask-dependent PIT criterion helps train the model in a more reliable way. Our future research will analyze how the mask-dependent PIT criterion can contribute to such end-to-end approaches.

Though we do not achieve better performance than \citep{wang2019deep}, it is good to mention the chimera++ network applied in \citep{wang2019deep} is different. It adds several convolutional encoder and decoder layers before and after the chimera++ network to achieve 11.9~dB SI-SDR ($\sim$12.2 to 12.4~dB SDR). Also, since the model applies the waveform approximate loss function to reconstruct the time-domain signal, similarly to Phasebook, we can adapt our proposed method to the model to improve the phase estimation accuracy further. \citep{liu2019divide} show that frame-level PIT criterion can find a better local minimum hence significantly improves the performance comparing than the utterance-level PIT. Our future plan is to integrate the frame-level PIT into our mask-dependent criterion and conduct the evaluation.

\section{Conclusion}
\label{sec:conc}
This paper proposed a phase estimation method using the mask estimate from the chimera++ network. A mask-dependent PIT criterion is applied to solve the label-permutation problem for both the mask and the phase estimates.
The mask-dependent PIT criterion significantly improved the separation performance compared with the PIT over the whole loss.
Future study will focus on applying the proposed PIT criterion to end-to-end phase estimation methods (e.g., phasebook \citep{le2019phasebook} and the sign prediction network \citep{wang2019deep}).

\section{ACKNOWLEDGMENTS}
\label{sec:ack}
This material is based upon work supported by a Google Faculty Research Award and the National Science Foundation (NSF) under Grant IIS-1409431.  Any  opinions, findings, and conclusions or recommendations expressed in this material are those of the author(s) and do not necessarily reflect the views of the NSF.

\bibliographystyle{IEEEbib}
\bibliography{refs19}

\begin{thebibliography}{10}

\bibitem{hershey2016deep}
John~R Hershey, Zhuo Chen, Jonathan Le~Roux, and Shinji Watanabe,
\newblock ``Deep clustering: Discriminative embeddings for segmentation and
  separation,''
\newblock in {\em 2016 IEEE International Conference on Acoustics, Speech and
  Signal Processing (ICASSP)}. IEEE, 2016, pp. 31--35.

\bibitem{erdogan2015phase}
Hakan Erdogan, John~R Hershey, Shinji Watanabe, and Jonathan Le~Roux,
\newblock ``Phase-sensitive and recognition-boosted speech separation using
  deep recurrent neural networks,''
\newblock in {\em 2015 IEEE International Conference on Acoustics, Speech and
  Signal Processing (ICASSP)}. IEEE, 2015, pp. 708--712.

\bibitem{yu2017permutation}
Dong Yu, Morten Kolb{\ae}k, Zheng-Hua Tan, and Jesper Jensen,
\newblock ``Permutation invariant training of deep models for
  speaker-independent multi-talker speech separation,''
\newblock in {\em 2017 IEEE International Conference on Acoustics, Speech and
  Signal Processing (ICASSP)}. IEEE, 2017, pp. 241--245.

\bibitem{luo2017deep}
Yi~Luo, Zhuo Chen, John~R Hershey, Jonathan Le~Roux, and Nima Mesgarani,
\newblock ``Deep clustering and conventional networks for music separation:
  Stronger together,''
\newblock in {\em 2017 IEEE International Conference on Acoustics, Speech and
  Signal Processing (ICASSP)}. IEEE, 2017, pp. 61--65.

\bibitem{wang2018alternative}
Zhong-Qiu Wang, Jonathan Le~Roux, and John~R Hershey,
\newblock ``Alternative objective functions for deep clustering,''
\newblock in {\em 2018 IEEE International Conference on Acoustics, Speech and
  Signal Processing (ICASSP)}. IEEE, 2018, pp. 686--690.

\bibitem{williamson2016complex}
Donald~S Williamson, Yuxuan Wang, and DeLiang Wang,
\newblock ``Complex ratio masking for monaural speech separation,''
\newblock {\em IEEE/ACM Transactions on Audio, Speech and Language Processing
  (TASLP)}, vol. 24, no. 3, pp. 483--492, 2016.

\bibitem{williamson2017speech}
Donald~S Williamson and DeLiang Wang,
\newblock ``Speech dereverberation and denoising using complex ratio masks,''
\newblock in {\em 2017 IEEE International Conference on Acoustics, Speech and
  Signal Processing (ICASSP)}. IEEE, 2017, pp. 5590--5594.

\bibitem{gunawan2010iterative}
David Gunawan and Deep Sen,
\newblock ``Iterative phase estimation for the synthesis of separated sources
  from single-channel mixtures,''
\newblock {\em IEEE Signal Processing Letters}, vol. 17, no. 5, pp. 421--424,
  2010.

\bibitem{wang2018end}
Zhong-Qiu Wang, Jonathan~Le Roux, DeLiang Wang, and John~R Hershey,
\newblock ``End-to-end speech separation with unfolded iterative phase
  reconstruction,''
\newblock {\em arXiv preprint arXiv:1804.10204}, 2018.

\bibitem{ephrat2018looking}
Ariel Ephrat, Inbar Mosseri, Oran Lang, Tali Dekel, Kevin Wilson, Avinatan
  Hassidim, William~T Freeman, and Michael Rubinstein,
\newblock ``Looking to listen at the cocktail party: A speaker-independent
  audio-visual model for speech separation,''
\newblock {\em arXiv preprint arXiv:1804.03619}, 2018.

\bibitem{afouras2018conversation}
Triantafyllos Afouras, Joon~Son Chung, and Andrew Zisserman,
\newblock ``The conversation: Deep audio-visual speech enhancement,''
\newblock {\em arXiv preprint arXiv:1804.04121}, 2018.

\bibitem{he2016deep}
Kaiming He, Xiangyu Zhang, Shaoqing Ren, and Jian Sun,
\newblock ``Deep residual learning for image recognition,''
\newblock in {\em Proceedings of the IEEE conference on computer vision and
  pattern recognition}, 2016, pp. 770--778.

\bibitem{paszke2017automatic}
Adam Paszke, Sam Gross, Soumith Chintala, Gregory Chanan, Edward Yang, Zachary
  DeVito, Zeming Lin, Alban Desmaison, Luca Antiga, and Adam Lerer,
\newblock ``Automatic differentiation in pytorch,''
\newblock in {\em NIPS-W}, 2017.

\bibitem{mcfee2015librosa}
Brian McFee, Colin Raffel, Dawen Liang, Daniel~PW Ellis, Matt McVicar, Eric
  Battenberg, and Oriol Nieto,
\newblock ``librosa: Audio and music signal analysis in python,''
\newblock in {\em Proceedings of the 14th python in science conference}, 2015,
  pp. 18--25.

\bibitem{raffel2014mir_eval}
Colin Raffel, Brian McFee, Eric~J Humphrey, Justin Salamon, Oriol Nieto, Dawen
  Liang, Daniel~PW Ellis, and C~Colin Raffel,
\newblock ``mir\_eval: A transparent implementation of common mir metrics,''
\newblock in {\em In Proceedings of the 15th International Society for Music
  Information Retrieval Conference, ISMIR}. Citeseer, 2014.

\bibitem{le2019phasebook}
Jonathan Le~Roux, Gordon Wichern, Shinji Watanabe, Andy Sarroff, and John~R
  Hershey,
\newblock ``Phasebook and friends: Leveraging discrete representations for
  source separation,''
\newblock {\em IEEE Journal of Selected Topics in Signal Processing}, 2019.

\bibitem{wang2019deep}
Zhong-Qiu Wang, Ke~Tan, and DeLiang Wang,
\newblock ``Deep learning based phase reconstruction for speaker separation: A
  trigonometric perspective,''
\newblock in {\em ICASSP 2019-2019 IEEE International Conference on Acoustics,
  Speech and Signal Processing (ICASSP)}. IEEE, 2019, pp. 71--75.

\bibitem{le2018sdr}
Jonathan Le~Roux, JR~Hershey, A~Liutkus, F~St{\"o}ter, ST~Wisdom, and
  H~Erdogan,
\newblock ``Sdr--half-baked or well done?,''
\newblock {\em Mitsubishi Electric Research Laboratories (MERL), Cambridge, MA,
  USA, Tech. Rep}, 2018.

\bibitem{isik2016single}
Yusuf Isik, Jonathan~Le Roux, Zhuo Chen, Shinji Watanabe, and John~R Hershey,
\newblock ``Single-channel multi-speaker separation using deep clustering,''
\newblock {\em arXiv preprint arXiv:1607.02173}, 2016.

\bibitem{bengio2009curriculum}
Yoshua Bengio, J{\'e}r{\^o}me Louradour, Ronan Collobert, and Jason Weston,
\newblock ``Curriculum learning,''
\newblock in {\em Proceedings of the 26th annual international conference on
  machine learning}. ACM, 2009, pp. 41--48.

\bibitem{liu2019divide}
Yuzhou Liu and DeLiang Wang,
\newblock ``Divide and conquer: A deep casa approach to talker-independent
  monaural speaker separation,''
\newblock {\em arXiv preprint arXiv:1904.11148}, 2019.

\end{thebibliography}
%
%
%
%
%
%
%
%
%

\end{sloppy}
\end{document}